\documentclass[twocolumn]{autart}    

\usepackage{graphicx}          
\usepackage{cite}
\usepackage{wrapfig,comment}
\usepackage{amsmath,amssymb,amsfonts,mathtools,textgreek}
\usepackage{algorithmic,enumerate}
\usepackage[dvipsnames]{xcolor}
\usepackage{textcomp}
\usepackage[normalem]{ulem}

\usepackage{marginnote,zref-savepos}
\newcounter{labelnote}
\makeatletter
\let\oldmarginnote\marginnote
\renewcommand*{\marginnote}[1]{%
 \begingroup\strut
  \stepcounter{labelnote}\zsaveposx {marginnote-\thelabelnote}
     \ifnum 0\zposx{marginnote-\thelabelnote}<1900000
      \reversemarginpar
      \oldmarginnote{\color{blue}#1}%
     \else
      \normalmarginpar
      \oldmarginnote{\color{blue}#1}%
     \fi
 \endgroup%
}
\makeatother

\begin{document}
\newtheorem{lemma}{Lemma}
\newtheorem{theorem}{Theorem}
\newtheorem{definition}{Definition}
\newtheorem{assumption}{Assumption}
\newtheorem{corollary}{Corollary}
\newtheorem{remark}{Remark}
\newtheorem{conjecture}{Conjecture}

\newcommand{\emily}[1]{\textcolor{red}{#1}}

\begin{frontmatter}
\title{Information-theoretic multi-time-scale partially observable systems with inspiration from leukemia treatment\thanksref{footnoteinfo}} \vspace{-7mm}

\thanks[footnoteinfo]{This paper was not presented at any conference. This work was completed while E. Jensen was a postdoctoral researcher at Northeastern University. Corresponding author: M. P. Chapman (see email address below).}
\author[ChapmanAddress]{Margaret P. Chapman}\ead{mchapman@ece.utoronto.ca},
\author[JensenAddress]{Emily Jensen}\ead{ejensen@colorado.edu},      
\author[ChanAddress]{Steven M. Chan}\ead{steven.chan@uhnresearch.ca},
\author[LessardAddress]{Laurent Lessard}\ead{l.lessard@northeastern.edu}

\address[ChapmanAddress]{Edward S. Rogers Sr. Department of Electrical and Computer Engineering, University of Toronto, 10 King's College Road, Toronto, ON, M5S 3G8, Canada}                   
\address[JensenAddress]{Department of Electrical, Computer and Energy Engineering, University of Colorado Boulder, 425 UCB, Boulder, CO, 80309,United States} 
\address[ChanAddress]{Department of Medical Biophysics, Faculty of Medicine, University of Toronto, Princess Margaret Cancer Research Tower, MaRS Centre, 101 College Street, Toronto, ON, M5G 1L7, Canada}
\address[LessardAddress]{Department of Mechanical and Industrial Engineering, Northeastern University, 334 Snell Engineering Center, 360 Huntington Avenue, Boston, MA, 02115, United States}
\vspace{-5mm}
          

\begin{abstract} 
We study a partially observable nonlinear stochastic system with unknown parameters, where the given time scales of the states and measurements may be distinct. The proposed setting is inspired by disease management, particularly leukemia. \vspace{-5mm} 
\end{abstract}
\end{frontmatter}

\section{Introduction}\label{Introduction}\vspace{-3mm}
A patient with a disease may be viewed as a stochastic system that evolves in response to control inputs, e.g., drug doses. Aspects of the patient can be measured, e.g., a tumor can be scanned or a blood sample can be taken. Controls and measurements typically operate on different time scales; for example, measurements may be weekly, and there may be days when no drug is taken. While there is knowledge about the underlying biochemical processes, such processes may be nonlinear, noisy, and may vary between patients. In many medical settings (e.g., leukemia treatment, diabetes treatment, and blood pressure management), the doses of drugs require adjustments for the purpose of regulation (e.g., keeping white blood cells, glucose, and blood pressure, respectively, within specific ranges). From a mathematical point of view, it is unclear how to unify the above properties (partial observability, nonlinear stochastic dynamics with uncertain parameters, and multiple time scales) into a rigorous control-theoretic framework. Our aim is to develop such a unifying framework. \vspace{-2mm}

We consider a discrete-time partially observable stochastic system that differs from a standard set-up. The system has an unknown deterministic parameter vector $\theta$, representing interpatient variability. Furthermore, the states and measurements may operate on different time scales, the dynamics and measurement equations may be nonlinear, the process and measurement noise may 
have unbounded support, and 
the spaces of states, controls, and measurements do not have finite cardinality.
Working in the setting above, we aim to develop a pathway that provides a control policy with awareness about the uncertainty in an estimate for $\theta$ and sensitivity of the state relative to $\theta$. \emph{A theoretical foundation must be developed before approximations can be investigated to permit computation}.\vspace{-3mm} 

\begin{figure}[b]
   \begin{center}
    \includegraphics[width = \columnwidth]{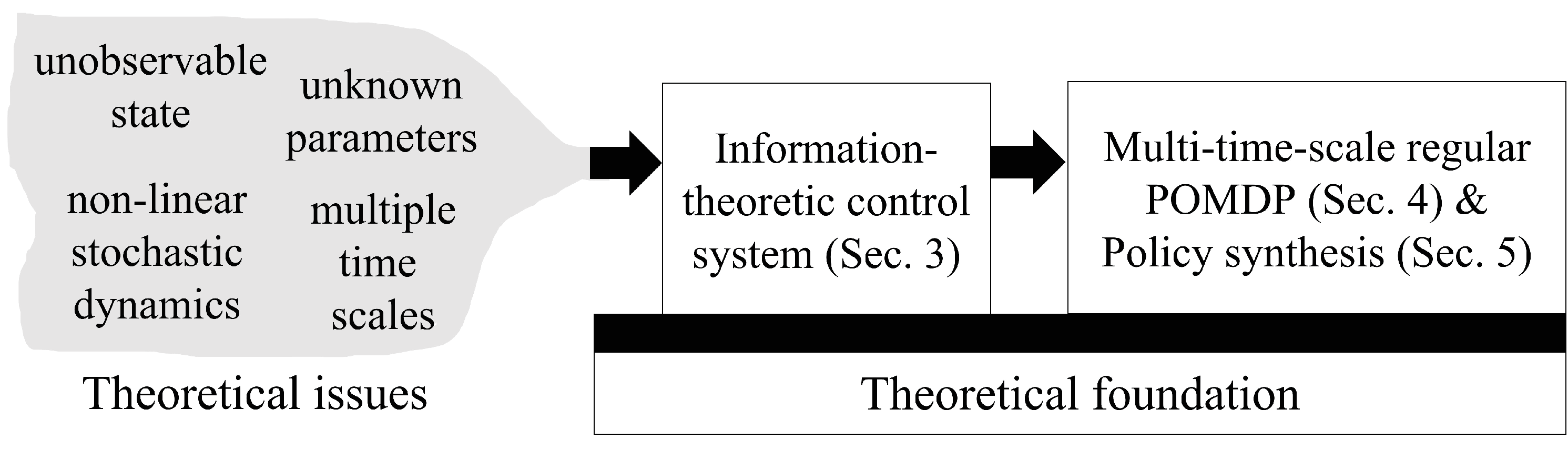}\vspace{-2mm}
    \caption{A high-level illustration of this work.}
    \label{keyidea}
\end{center}
\end{figure}

Our contribution is of a conceptual nature. We offer a theoretical framework that unifies the following key characteristics: partial observability, uncertain parameters, multiple time scales, and nonlinear stochastic dynamics (Fig. \ref{keyidea}). First, we devise a theoretical approach to handle the unknown parameters, building on concepts from optimum experiment design and dual control (Sec. \ref{secII}). Second, we represent the system as a multi-time-scale partially observable (PO) Markov decision process (MDP) with information-theoretic cost functions and show that it enjoys regularity properties under some assumptions (Sec. \ref{secIII}). Third, we provide conditions that guarantee the existence of an optimal policy 
for a belief-space MDP corresponding to the POMDP 
(Sec. \ref{secexistence}). 
Our main contribution is to reduce the complications that arise from the unobservable state, unknown parameters, multiple time scales, and nonlinear stochastic dynamics to a form that admits a mathematical solution. We exemplify the assumptions in the context of a leukemia treatment model in Section \ref{example}. 
Although our presented framework does not solve the curse of dimensionality, it does provide a launching point for bridging the gap between theory and practice, as further described in our conclusion (Sec. 7).
 \vspace{-1mm}

\vspace{-2mm}

\subsection{Related literature}\vspace{-4mm}
The theory is related to three bodies of literature: POMDPs, optimum experiment design, and stochastic adaptive and dual control. 
POMDPs with Borel state and control spaces have been studied since at least the 1970s \cite{rhenius1974incomplete} \cite{yushkevich1976reduction} \cite[Ch. 10]{bertsekas2004stochastic} \cite[Ch. 4]{hernandez1989adaptive}. The state-of-the-art conditions that facilitate policy synthesis for these general POMDPs were developed recently \cite{feinberg2016partially} \cite{kara2019weak} \cite{feinberg2022markov}. We propose and study a POMDP problem with unknown parameters and information-theoretic costs, where the time scales of the measurements and states may be distinct. The different time scales in the setting of unknown parameters pose difficulties due to the irregular updates to the parameter estimates and posterior distributions. 
We offer a reformulation that uses measure-theoretic first principles and regularity conditions proposed by \cite{feinberg2016partially} \cite{kara2019weak}. The author of \cite[Ch. 4.5]{hernandez1989adaptive} studies POMDPs with unknown parameters, continuous bounded cost functions, and weakly continuous bounded dynamics for posterior distributions.
However, we consider cost functions that are lower semicontinuous and bounded below. We do not assume weakly continuous bounded dynamics for posterior 
distributions in view of a counterexample \cite[Ex. 4.2]{feinberg2016partially}. 
We propose stage costs that penalize poor quality of information from measurements using the \emph{Fisher information matrix} from optimum experiment design (to be discussed).
To accommodate these information-theoretic costs, we assume Euclidean spaces of states, parameters, measurements, and measurement noise, and we make assumptions about the existence and continuity of Jacobians. \vspace{-2mm}

An optimum experiment design (OED) problem is a type of optimal control problem \cite{sager2013sampling}. An example OED problem is to optimize the planned measurement times by maximizing the quality of information, which can be quantified using the Fisher information matrix (FIM) \cite{pronzato2008optimal} \cite{sager2013sampling} \cite{nimmegeers2020optimal} \cite{erdal2021optimal}. 
Then, an optimal control problem can be solved using the parameter estimate \cite[Eq. (3.3)]{sager2013sampling} \cite[Algorithm 2, pp. 79--81]{jost2020model}. 
In clinical practice, the planned measurement times are determined by the patient's schedule, which may be infeasible to adjust. Thus, we are not concerned with optimizing the planned measurement times. 
The OED literature includes measurement noise but often neglects process noise \cite{sager2013sampling} \cite{la2017dual} \cite{jost2020model} \cite{erdal2021optimal}; an exception is 
\cite{telen2013optimal}.
We include both measurement and process noise; the latter plays numerous roles in biochemical systems \cite{rao2002control} \cite{eling2019challenges}. 
The techniques in \cite{telen2013optimal} rely on linear approximations and a Riccati differential equation to approximate the covariance of the estimated state at the final time. In contrast, we leverage the theory of discrete-time nonlinear POMDPs for policy synthesis. Our objective evaluates the stage-wise performance of the states and controls and an FIM-based criterion. \vspace{-1.5mm}

This paper has connections to stochastic adaptive and dual control, which are related to OED. 
Separating the problems of parameter estimation and controller design characterizes classical adaptive control schemes, e.g., self-tuning regulators \cite[p. 22]{aastrom2013adaptive}.
In general, separating the two problems implies that parameter uncertainties are neglected in controller synthesis, 
and hence, dual control methods have been developed to tackle the two problems simultaneously \cite[pp. 22--24]{aastrom2013adaptive}.
A recent review of stochastic dual control is provided by \cite{mesbah2018stochastic}, and a concise summary is provided by \cite[p. 276]{hewing2020learning}.
We discuss some recent papers on dual control, which involve systems with unknown parameters (exception: \cite{feng2018real}).
Model predictive controllers for linear systems with process noise \cite{bavdekar2016stochastic} 
and nonlinear systems without process noise \cite{la2017dual} have been developed, where the objective is a sum of a performance metric and an FIM-based metric. 
The authors of \cite{la2017dual} also assume initial state distributions with bounded support.
Heirung \emph{et al.} develop a model predictive controller using a quadratically-constrained quadratic program to minimize the predicted mean-squared output error; the system is a linear regression model with normally distributed noise and parameters \cite{heirung2017dual}.
Model predictive controllers with information matrix-based constraints have been proposed for linear systems with process noise \cite{larsson2016application} and nonlinear systems without process noise \cite{telen2017study}. The controller in \cite{larsson2016application} incorporates different time scales for the measurements and controls.
Feng and Houska develop a real-time model predictive control algorithm for a nonlinear system
to optimize a sum of a performance metric and an approximation for the average loss of optimality due to 
poor future state estimates \cite{feng2018real}. They assume that the measurement and process noise have bounded support to facilitate 
the latter approximation, which employs an extended Kalman filter
\cite{feng2018real}.
In contrast to the above papers, we provide a theoretical study of a partially observable multi-time-scale nonlinear system with unknown parameters, where the measurement and process noise may have unbounded support. We include an FIM-based criterion in the objective rather than as a constraint to avoid potential feasibility issues. \vspace{-3mm}

\subsection{Notation}\label{notationsection}\vspace{-3.5mm}
$\mathbb{R}$ is the real line. $\mathbb{R}^* \coloneqq \mathbb{R} \cup \{-\infty,+\infty\}$ is the extended real line. $\mathbb{R}^n$ is $n$-dimensional Euclidean space, $\mathbb{R}_+^n \coloneqq \{x \in \mathbb{R}^n : x_i > 0, i = 1,2,\dots,n \}$, and $\mathbb{R}_+ \coloneqq \mathbb{R}_+^1$. $\mathbb{N}$ is the set of natural numbers. If $f : \mathbb{R}^n \times \mathbb{R}^l \rightarrow \mathbb{R}^m$ with $(x,y) \mapsto f(x,y)$, 
then $\frac{\partial f(a,b)}{\partial x} \in \mathbb{R}^{m \times n}$ is the Jacobian matrix of $f$ with respect to $x$ evaluated at $(a,b) \in \mathbb{R}^n \times \mathbb{R}^l$. 
The Euclidean norm is $\|\cdot \|$. If $A \in \mathbb{R}^{m \times n}$, we define $(*)^\top A \coloneqq A^\top A$.
\textcolor{black}{We define 
$\varphi_\epsilon : \mathbb{R}^l \rightarrow \mathbb{R}^l_+$ by
 $\varphi_\epsilon(x) \coloneqq (\max\{x_1,\epsilon\}, \max\{x_2,\epsilon\}, \dots, \max\{x_l,\epsilon\})^\top$, where
$\epsilon \in \mathbb{R}_+$ is near zero.}
If $M$ is a separable metrizable space, then $\mathcal{B}_M$ is the Borel sigma algebra on $M$, 
$\mathcal{P}_M$ is the collection of probability measures on $(M,\mathcal{B}_M)$ with the weak topology, 
and $\|\cdot \|_{\text{TV}}$ is the total variation norm on $\mathcal{P}_M$. If $p_i \in \mathcal{P}_M$ for $i = 1,2$, the total variation norm of $p_1 - p_2$ is $\|p_1 - p_2\|_{\text{TV}} = 2 \sup_{\underline{M} \in \mathcal{B}_M}| p_1(\underline{M}) - p_2(\underline{M}) |$.
We use the notation $\underline{M}$ to denote a Borel set, i.e., $\underline{M} \in \mathcal{B}_M$, following the notation from \cite{bertsekas2004stochastic}.
If $y \in M$, then $\delta_y$ is the Dirac measure in $\mathcal{P}_{M}$ concentrated at $y$.
If $A$ and $B$ are two sets, then $A \setminus B$ means $A$ set minus $B$.
Note the following abbreviations: l.s.c. = lower semicontinuous,
b.b. = bounded below, measurable = Borel-measurable, and Rmk. = Remark. \vspace{-2mm}

\subsection{Preliminaries}\vspace{-3.5mm}
To keep the work self-contained, we recall some principles mostly from \cite{bertsekas2004stochastic}. Related material is available from \cite{hinderer1970decision} \cite{ash1972probability} \cite{hernandez1989adaptive} \cite{hernandez2012discrete}. 
$\text{X}$ and $\text{Y}$ are separable metrizable spaces. 
\vspace{-5mm}
\begin{remark}[Weak convergence]\label{weakconvremark}
\emph{Let $(p_n)_{n \in \mathbb{N}}$ be a sequence in $\mathcal{P}_\text{X}$ and $p \in \mathcal{P}_\text{X}$. The (weak) convergence of $p_n \rightarrow p$ in $\mathcal{P}_\text{X}$ is equivalent to $\int_\text{X} f \;\mathrm{d}p_n \rightarrow \int_\text{X} f \; \mathrm{d}p$ in $\mathbb{R}$ for every continuous bounded function $f : \text{X} \rightarrow \mathbb{R}$ \cite[Prop. 7.21]{bertsekas2004stochastic}.
}
\end{remark}\vspace{-2mm}
\begin{remark}[Stochastic kernels]\label{def2}
\emph{A \emph{measurable\\stochastic kernel} $q(\mathrm{d}y|x)$ on $\text{Y}$ given $\text{X}$ is a family of elements of $\mathcal{P}_\text{Y}$ parametrized by elements of $\text{X}$, where the map $v: \text{X} \rightarrow \mathcal{P}_{\text{Y}}$ defined by $v(x) \coloneqq q(\cdot|x)$ is measurable \cite[Def. 7.12]{bertsekas2004stochastic}. If $v$ is weakly continuous, i.e., $x_n \rightarrow x$ in $\text{X}$ implies $q(\cdot|x_n) \rightarrow q(\cdot|x)$ in $\mathcal{P}_{\text{Y}}$, then $q$ is called \emph{weakly continuous}. If $v$ is continuous in total variation, i.e., $x_n \rightarrow x$ in $\text{X}$ implies $\|q(\cdot|x_n) - q(\cdot|x)\|_{\text{TV}} \rightarrow 0$, then $q$ is called \emph{continuous in total variation}.}
\end{remark}\vspace{-2mm}
\begin{remark}[Some continuity facts]\label{remark3}
\emph{
Let $q$ be a weakly continuous stochastic kernel on $\text{Y}$ given $\text{X}$, $f : \text{X} \times \text{Y} \rightarrow \mathbb{R}^*$ be measurable and b.b., and $g : \text{X} \rightarrow \mathbb{R}^*$ be defined by $g(x) \coloneqq \int_{\text{Y}} f(x,y) \; q(\mathrm{d}y|x)$. If $f$ is continuous and bounded, then 
$g$ is continuous and bounded \cite[Prop. 7.30]{bertsekas2004stochastic}. 
If $f$ is l.s.c. (and b.b.), then $g$ is l.s.c. and b.b. \cite[Prop. 7.31]{bertsekas2004stochastic}.
The map $\ell : \text{Y} \rightarrow \mathcal{P}_{\text{Y}}$ defined by $\ell(y) \coloneqq \delta_y$ is weakly continuous \cite[Cor. 7.21.1]{bertsekas2004stochastic}. 
The map $\sigma : \mathcal{P}_{\text{X}} \times  \mathcal{P}_{\text{Y}} \rightarrow \mathcal{P}_{\text{X} \times \text{Y}}$ defined by $\sigma(p_1,p_2) \coloneqq p_1 p_2$ is weakly continuous \cite[Lemma 7.12]{bertsekas2004stochastic}. 
The construction of the product $p_1 p_2 \in  \mathcal{P}_{\text{X} \times \text{Y}}$ of $p_1 \in \mathcal{P}_{\text{X}}$ and $p_2 \in \mathcal{P}_{\text{Y}}$ is provided by \cite[Cor. 2.6.3]{ash1972probability}.
}
\end{remark}\vspace{-2mm}
\begin{remark}[Kernel $\delta_{f}$]\label{remark4}
\emph{If $f : \text{X} \rightarrow \text{Y}$ is measurable, we define $\delta_{f}$ by $\delta_{f(x)} \coloneqq \ell(f(x))$ for all $x \in \text{X}$,
where $\ell$ is the Dirac measure map defined in Remark \ref{remark3}.
$\delta_{f}$ is a measurable stochastic kernel on $\text{Y}$ given $\text{X}$
due to the measurability of $f$ and the (weak) continuity of $\ell$.
}
\end{remark}\vspace{-2mm}
\begin{remark}[Measurable selection]\label{measselection}
\emph{Assume that $\text{Y}$ is compact. Let $f : \text{X} \times \text{Y} \rightarrow \mathbb{R}^*$ be l.s.c. and b.b., and define $f^* : \text{X} \rightarrow \mathbb{R}^*$ by $f^*(x) \coloneqq \inf_{y \in \text{Y}} f(x,y)$. Then, $f^*$ is l.s.c. and b.b., and there exists a measurable function $\kappa : \text{X} \rightarrow \text{Y}$ such that $f(x,\kappa(x)) = f^*(x)$ for all $x \in \text{X}$ \cite[Prop. 7.33]{bertsekas2004stochastic}.} 
\end{remark}\vspace{-2mm}

\section{System description}\label{sysdescription}\vspace{-3mm}
We consider a discrete-time stochastic system that 
differs from a standard one. There is an unknown deterministic parameter vector, there may be different time scales, and the state is not observable.
The system takes the following form:\vspace{-2mm}
\begin{equation}\label{gensys}
\begin{aligned}
    x_{t+1} & = f_t(x_t,u_t,d_t; \theta), && t \in \mathbb{T}, \\
    y_t & = h_t(x_t; \theta) + w_t, && t \in \mathbb{T}_y, 
\end{aligned}\vspace{-2mm}
\end{equation}
where $x_t \in \mathbb{S}$ is a state, $y_t \in \mathbb{Y} \coloneqq \mathbb{R}^m$ is a measurement, $u_t \in \mathbb{C}$ is a control, $d_t \in \mathbb{D}$ is a process noise realization, $w_t \in \mathbb{Y}$ is a measurement noise realization, and $\theta \in \mathbb{P} \coloneqq \mathbb{R}_+^p$ is a parameter vector whose true value is unknown. An initial estimate for $\theta$ is available. 
The state space $\mathbb{S} \in \mathcal{B}_{\mathbb{R}^n}$
is a nonempty Euclidean
space. The control space $\mathbb{C}$ and the process noise space $\mathbb{D}$ are nonempty Borel spaces \cite[Def. 7.7]{bertsekas2004stochastic}.
The states evolve on $\mathbb{T} \coloneqq \{0,1,\dots,N-1\}$, a time horizon of length $N \in \mathbb{N}$. The measurements evolve on a nonempty subset $\mathbb{T}_y$ of $\mathbb{T}_N \coloneqq \mathbb{T} \cup \{N\}$. 
The controls may be optimized on a nonempty subset $\mathbb{T}_u$ of $\mathbb{T}$. If $t \in \mathbb{T} \setminus \mathbb{T}_u$, then $u_t$ is assigned a given default value $\textbf{u} \in \mathbb{C}$.
The functions $f_t : \mathbb{S} \times \mathbb{C} \times \mathbb{D} \times \mathbb{P} \rightarrow \mathbb{S}$ for every $t \in \mathbb{T}$ and $h_t : \mathbb{S} \times \mathbb{P} \rightarrow \mathbb{Y}$ for every $t \in \mathbb{T}_y$ are measurable. These functions and the horizons $\mathbb{T}$, $\mathbb{T}_y$, and $\mathbb{T}_u$ are given.
The quantities $x_t$, $u_t$, $d_t$, $y_t$, and $w_t$ in \eqref{gensys} are realizations of random objects $X_t$, $U_t$, $D_t$, $Y_t$, and $W_t$, respectively.
The random objects $X_0$, $D_t$ for every $t \in \mathbb{T}$, and $W_t$ for every $t \in \mathbb{T}_y$ are independent. $D_t$ for any $t \in \mathbb{T}$ and $W_t$ for any $t \in \mathbb{T}_y$ do not depend on $\theta$.
The (prior) distribution of $X_0$
is given, and the distributions of $D_t$ and $W_t$ are given when these random objects are defined. (Assuming knowledge of such distributions is typical in research about partially observable systems.) We use the following notations:\vspace{-2mm}
\begin{itemize}
    \item If $t \in \mathbb{T} \setminus \mathbb{T}_y$, then the distribution of $D_t$ is $\mu_t \in \mathcal{P}_{\mathbb{D}}$.
    \item If $t \in \mathbb{T}_y$, then the distribution of $W_t$ is $\rho_t \in \mathcal{P}_{\mathbb{Y}}$.
    \item If $t \in \mathbb{T}_y \setminus \{N\}$, then the distribution of $(D_t,W_t)$ is $\nu_{t} \in \mathcal{P}_{\mathbb{D} \times \mathbb{Y}}$.\vspace{-2mm}
\end{itemize}
$\rho_t$ and $\nu_t$ are related by
    $\rho_t(\underline{\mathbb{Y}}) = \nu_t(\mathbb{D} \times \underline{\mathbb{Y}})$ for every $\underline{\mathbb{Y}} \in \mathcal{B}_{\mathbb{Y}}$ and $t \in \mathbb{T}_y\setminus \{N\}$.
The system \eqref{gensys} has a stage cost function $\hat{c}_t : \mathbb{S} \times \mathbb{C} \times \mathbb{P} \rightarrow \mathbb{R}^*$ for every $t \in \mathbb{T}$ and a terminal cost function $\hat{c}_N : \mathbb{S} \times \mathbb{P} \rightarrow \mathbb{R}^*$, which may depend on $\theta$. These functions assess the performance of the states and controls. We invoke the following assumption. \vspace{-1.5mm} 
\begin{assumption}[About $f_t$, $\hat{c}_t$, $\mathbb{C}$, $h_t$, $\rho_t$]\label{assumption1}
\emph{We assume:
\vspace{-3mm}\begin{enumerate}[(a)]
    \item $f_t$ is continuous for every $t \in \mathbb{T}$;
    \item $\hat{c}_t$ is l.s.c. and bounded below for every $t \in \mathbb{T}_N$;
    \item $\mathbb{C}$ is compact;
    \item $h_t$ is continuous for every $t \in \mathbb{T}_y$;
    \item $\rho_t$ admits a continuous (nonnegative) density with respect to Lebesgue measure in $\mathcal{P}_{\mathbb{Y}}$ for every $t \in \mathbb{T}_y$;
    \item $f_t(x,u,d;\theta)$ is differentiable in $x$ and $\theta$ for every $t \in \mathbb{T}$; $h_t(x;\theta)$ is differentiable in $x$ and $\theta$ for every $t \in \mathbb{T}_y$; the associated Jacobians are continuous.
\end{enumerate}\vspace{-3mm}}
\end{assumption}
We abbreviate Assumption \ref{assumption1} as A1.
Parts (a)--(c) of A1 are standard conditions to help ensure the existence of an optimal policy when the objective is an expected cumulative cost and the state is observable \cite[Def. 8.7]{bertsekas2004stochastic}.
Parts (d)--(e) will help guarantee the regularity of observation kernels.
Part (f) will facilitate estimating $\theta$ using measurements and 
assessing the quality of the measurements. In particular, our leukemia treatment model satisfies A1 (Sec. \ref{example}).\vspace{-3mm}
\section{Information-theoretic control system}\label{secII}\vspace{-3mm}

Toward designing a control policy for \eqref{gensys}, here we provide a pathway for estimating $\theta$ and managing potential differences between $\mathbb{T}_N$ and $\mathbb{T}_y$.
First, we specify cost functions to assess the quality of measurements using concepts from optimum experiment design. Second, we define an enlarged system to record these information-theoretic costs. We conclude the section by studying the system's properties in Theorem \ref{thm1}. \vspace{-1.5mm}

\subsection{Formulating information-theoretic costs}\vspace{-3mm}
We would like to extract useful information from measurements to inform the estimation of $\theta$. The OED literature assesses the usefulness of information through the Fisher information matrix (FIM). For the system \eqref{gensys} and a state trajectory $(x_0,x_1,\dots,x_{N}) \in \mathbb{S}^{N+1}$, the FIM $\mathcal{F}\in \mathbb{R}^{p \times p}$ is defined by $\mathcal{F}\coloneqq \sum_{t \in \mathbb{T}_y} \mathcal{F}_t((x_t^\top, \xi_t^\top, \theta^\top)^\top)$ \cite[Def. 3.2]{sager2013sampling}, 
where each term \vspace{-3mm}
\begin{subequations}\label{fishert}
\begin{align}
     \mathcal{F}_t\left((x_t^\top, \xi_t^\top, \theta^\top)^\top\right)
    & \coloneqq (*)^\top \left( \frac{\mathrm{d}h_t(x_t;\theta)}{\mathrm{d}\theta} \right), \quad t \in \mathbb{T}_y,   \vspace{-6mm}
\end{align}
depends on the \emph{output sensitivity} matrix \cite[Eq. (6.101)]{keesman2011system} 
\begin{equation}\label{outputsen}
    \frac{\mathrm{d}h_t(x_t;\theta)}{\mathrm{d}\theta} \coloneqq \frac{\partial h_t(x_t;\theta)}{\partial x} \cdot \xi_t + \frac{\partial h_t(x_t;\theta)}{\partial \theta}, \quad t \in \mathbb{T}_y,
\end{equation}
\end{subequations}
and the \emph{state sensitivity} matrix $\xi_t \coloneqq \frac{\mathrm{d}x_t}{\mathrm{d} \theta} \in \mathbb{R}^{n \times p}$.
The output sensitivity matrix  
\eqref{outputsen}
is the total derivative of the predicted measurement $h_t(x_t;\theta)$ with respect to $\theta$. 
The $\xi_t$-dynamics are defined by\footnote{
The definition \eqref{xidyn} uses the mild assumption that the process noise does not depend on $\theta$. 
Moreover, the definition \eqref{xidyn} neglects the dependency of $u_t$ on $\theta$. This simplification is standard, e.g., see \cite[Def. 3.1]{sager2013sampling} 
and \cite[Eq. (6.100)]{keesman2011system} for continuous-time examples, because how $u_t$ depends on the information available up to time $t$, i.e., $i_t \coloneqq (y_0,u_0,\dots,y_{t-1},u_{t-1},y_t)$,
is not known \emph{a priori}. An alternative is to restrict oneself to controls of the form $u_t = K_t(i_t)$, where $K_t$ is a linear combination of differentiable functions, e.g., polynomials, so that $\frac{\mathrm{d}u_t}{\mathrm{d}\theta}$ can be evaluated and included in the $\xi_t$-dynamics \eqref{xidyn}. While this alternative is out of scope, it may be interesting for future study.
}\vspace{-1mm}
\begin{subequations}\label{xidyn}
\begin{equation}
    \xi_{t+1} = \phi_t(\xi_t,x_t,u_t,d_t; \theta), \quad t \in \mathbb{T},\vspace{-1.5mm}
\end{equation}
where\vspace{-1.5mm}
\begin{equation}\label{myphit}
    \phi_t(\xi,x,u,d; \theta) \coloneqq \frac{\partial f_t(x,u,d;\theta)}{\partial x} \cdot \xi + \frac{\partial f_t(x,u,d;\theta)}{\partial \theta}.
\end{equation}
\end{subequations}
If $\theta$ does not include the initial state $x_0$, then $\xi_0$ is the zero matrix in $\mathbb{R}^{n \times p}$. If $\theta$ does include the initial state, then $\xi_0$ consists of an identity matrix and a zero matrix. \vspace{-1.5mm}

We would like the size of the FIM $\mathcal{F}$, e.g., its trace or determinant, to be large \cite[Def. 3.4]{sager2013sampling}. 
Statistical background underlying $\mathcal{F}$ can be found in \cite{ljung1999system}. We content ourselves with an intuitive explanation. A large $\text{trace}(\mathcal{F})$ means that the predicted measurement $h_t(x_t;\theta)$ varies a large amount with respect to $\theta$. Therefore, different values of $\theta$ correspond to different measurements, which facilitates the estimation of the true $\theta$ from the measurements.
Defining an augmented state that includes the state sensitivity matrix $\xi_t$ is useful for assessing FIM-based criteria in optimal control problems \cite{sager2013sampling} \cite{larsson2016application}. \vspace{-1.5mm}

The equations \eqref{fishert}--\eqref{xidyn} depend on $\theta$, whose true value is unknown. Hence, one evaluates these equations using an estimate for $\theta$, which we denote by 
$\hat{\theta}_t \in \mathbb{P}$. Consequently, we must track the evolution of the estimates, which is an aspect of dual control
\cite[p. 23]{aastrom2013adaptive} 
\cite[Ch. 2.5]{hernandez1989adaptive}. \vspace{-1.5mm}

In view of the above, we consider an augmented state $\chi_t \coloneqq (x_t^\top,\xi_t^\top,\hat{\theta}_t^\top)^\top$ with values in $\mathbb{X} \coloneqq \mathbb{S} \times \mathbb{R}^{n \times p} \times \mathbb{P}$.
We specify a 
cost function $\bar{c}_t : \mathbb{X} \rightarrow \mathbb{R}$, which aims to penalize a small trace of $\mathcal{F}_t(\chi)$ \eqref{fishert} for every $\chi = (x^\top,\xi^\top,\hat{\theta}^\top)^\top \in \mathbb{X}$, \vspace{-2mm}
\begin{equation}\label{14}
    \bar{c}_t(\chi) \coloneqq -\min\{\text{trace}(\mathcal{F}_t(\chi)), b \}, \quad t \in \mathbb{T}_y, \vspace{-2mm}
\end{equation}
where $b \in \mathbb{R}_+$ is a large constant. One selects $b$ empirically so that $\min\{\text{trace}(\mathcal{F}_t(\chi)), b \} = \text{trace}(\mathcal{F}_t(\chi))$ for all $\chi \in \mathbb{X}$ of practical interest. $b$ is needed so that $\bar{c}_t$ is b.b.\vspace{-1.5mm} 

Using $\bar{c}_t$ \eqref{14} and $\hat{c}_t$ (Sec. \ref{sysdescription}), we define stage cost functions $c_t : \mathbb{X} \times \mathbb{C} \rightarrow \mathbb{R}^*$ for all $t \in \mathbb{T}$ that include both information-theoretic and performance-based criteria: \vspace{-1.5mm}
\begin{subequations}\label{finalcosts}
\begin{equation}
\begin{aligned}
    c_t(\chi,u) & \coloneqq \begin{cases} \hat{c}_t(x,u; \hat{\theta}) + \lambda \bar{c}_t(\chi), & \text{if }t \in \mathbb{T}_y \setminus \{N\}, \\ \hat{c}_t(x,u; \hat{\theta}), & \text{if }t \in \mathbb{T} \setminus \mathbb{T}_y. \end{cases} \vspace{-2mm}
\end{aligned}
\end{equation}
$\lambda \in \mathbb{R}_+$ 
is chosen \emph{a priori} based on the relative importance of the two types of criteria. Similarly, we define a terminal cost function $c_N : \mathbb{X} \rightarrow \mathbb{R}^*$ by\vspace{-1.5mm}
\begin{equation}
       c_N(\chi) \coloneqq \begin{cases} \hat{c}_N(x; \hat{\theta}) + \lambda \bar{c}_N(\chi), & \text{if }N \in \mathbb{T}_y, \\ \hat{c}_N(x; \hat{\theta}), & \text{if }N \notin \mathbb{T}_y. \end{cases} \vspace{-1.5mm}
\end{equation}
\end{subequations}
A terminal cost is important for leukemia treatment because oncologists specify that the concentration of neutrophils should be within particular bounds by the end of a treatment cycle (to be exemplified in Sec. \ref{example}).  \vspace{-1.5mm}

An advantage of incorporating the FIM into a cost function, as in \eqref{finalcosts}, is that this choice avoids feasibility issues, which may arise if the FIM is incorporated into a constraint. However, $\lambda$ requires tuning. The authors of \cite{feng2018real} propose a trade-off term so that their objective penalizes an expected loss of optimality, if the noise has sufficiently small bounded support \cite[Th. 1]{feng2018real}. In contrast, we permit unbounded noise.
To formulate an optimal control problem using the cost functions \eqref{finalcosts}, 
we will define the dynamics of $\chi_t$ in the next subsection, starting with 
the dynamics of the parameter estimate $\hat{\theta}_t$. \vspace{-1.5mm}

\subsection{Defining $\chi_t$-dynamics}\vspace{-3mm}
We define the dynamics of $\hat{\theta}_t \in \mathbb{P}$ for every $t \in \mathbb{T}$ by\vspace{-3mm}
\begin{equation}\label{15a}
    \hat{\theta}_{t+1} = g_t(y_t,x_t;\hat{\theta}_t) \coloneqq \begin{cases} \bar{g}_t(y_t,x_t;\hat{\theta}_t), & \text{if }t \in \mathbb{T}_y \setminus \{N\}, \\
    \hat{\theta}_t, & \text{if } t \in \mathbb{T} \setminus \mathbb{T}_y, \end{cases}
\end{equation}
where 
$\hat{\theta}_0$ is an initial guess for $\theta$. For every $t \in \mathbb{T}_y \setminus \{N\}$, we define $\bar{g}_t$ using a gradient-based procedure \cite[Eq. (5.73)]{keesman2011system} 
adapted to ensure that $\bar{g}_t$ is $\mathbb{P}$-valued:
\vspace{-1.5mm} 
\begin{align}\label{17}
    \bar{g}_t(y,x;\theta) \coloneqq \varphi_\epsilon\big(\theta - \alpha_t \cdot L_t(x; \theta)^{-1} \cdot \text{D}_\theta \varepsilon_t(y,x;\theta)\big).
    \vspace{-2.5mm}
\end{align}
In \eqref{17}, $\varphi_\epsilon : \mathbb{R}^p \rightarrow \mathbb{P}$ is continuous (Sec. \ref{notationsection}), $\alpha_t \in \mathbb{R}_+$ is a step size,
$\text{D}_\theta \varepsilon_t(y,x;\theta) \in \mathbb{R}^p$ is the gradient of the mean-squared measurement errors 
$\varepsilon_t(y,x;\theta) \coloneqq \| y - h_t(x;\theta) \|^2$ 
with respect to $\theta$, and
$L_t(x; \theta) \in \mathbb{R}^{p \times p}$ is symmetric positive definite (to be specified).
We may choose $L_t(x; \theta)$ to be the identity matrix $I_{p}$ so that \eqref{17} resembles gradient descent. Otherwise, we use a Gauss-Newton update with a regularization term $\gamma \in \mathbb{R}_+$ \cite[Eq. (5.79)]{keesman2011system}: \vspace{-.5mm}
\begin{equation}\label{myRt}
    L_t(x; \theta) = 2  (*)^\top \left(\frac{\partial h_t(x;\theta)}{\partial \theta}\right) + \gamma I_{p}.   \vspace{-4mm}
\end{equation}

The $\chi_t$-dynamics depend on an output equation $H_t$ and a state update equation $F_t$. 
Using $h_t$ \eqref{gensys} and an arbitrary vector $\text{v} \in \mathbb{Y}$, 
we define $H_t : \mathbb{X} \times \mathbb{Y} \rightarrow \mathbb{Y}$ 
by \vspace{-3mm}
\begin{equation}\label{Ht}
    H_t(\chi,w) \coloneqq \begin{cases} h_t(x; \hat{\theta}) + w, & \text{if } t \in \mathbb{T}_y, \\ \text{v}, & \text{if } t \in \mathbb{T}_N \setminus \mathbb{T}_y. \end{cases}\vspace{-3mm}
\end{equation}
$\text{v}$ is a ``dummy'' vector that will be useful for writing the $\chi_t$-dynamics concisely and for analyzing these dynamics. We define $F_t : \mathbb{X} \times \mathbb{C} \times \mathbb{D} \times \mathbb{Y} \rightarrow \mathbb{X}$ by\vspace{-1.5mm}
\begin{equation}\label{mynewFta}
    F_t(\chi,u,d,w) \coloneqq \bar{F}_t(\chi,u,d,H_t(\chi,w)), \quad t \in \mathbb{T},\vspace{-1.5mm}
\end{equation}
where $\bar{F}_t$ depends on $f_t$ \eqref{gensys}, $\phi_t$ \eqref{xidyn}, and $g_t$ \eqref{15a} as follows: \vspace{-1.5mm}
\begin{equation}\label{18c}
    \bar{F}_t\left(\begin{bmatrix} x\\\xi\\\hat{\theta} \end{bmatrix},u,d,y\right) \coloneqq \begin{bmatrix} f_t(x,u,d;\hat{\theta}) \\ \phi_t(\xi,x,u,d;\hat{\theta}) \\ g_t(y,x;\hat{\theta}) \end{bmatrix}, \quad t \in \mathbb{T}.\vspace{-1.5mm}
\end{equation}
Then, the $\chi_t$-dynamics are given by\vspace{-1mm}
\begin{equation}\label{augsys}
\begin{aligned}
    \chi_{t+1} & = F_t(\chi_t,u_t,d_t,w_t), && t \in \mathbb{T},\\
    y_t & = H_t(\chi_t,w_t), && t \in \mathbb{T}_N, \vspace{-4mm}
\end{aligned}
\end{equation}
where $w_t$ is a realization of $W_t$ if $t \in \mathbb{T}_y$, and $w_t = \text{v}$ if $t \in \mathbb{T}_N \setminus \mathbb{T}_y$. 
$\text{v}$ does not affect $\chi_{t+1}$ for any $t \in \mathbb{T}$.
For every $t \in \mathbb{T} \setminus \mathbb{T}_y$ and $(\chi,u,d,w) \in \mathbb{X} \times \mathbb{C} \times \mathbb{D} \times \mathbb{Y}$, we have that $F_t(\chi,u,d,w) =  G_t(\chi,u,d)$, where \vspace{-1.5mm}
\begin{equation}\label{myGt}
 G_t(\chi,u,d) \hspace{-.3mm} \coloneqq \hspace{-.3mm} \begin{bmatrix} f_t(x,u,d;\hat{\theta}) \\ \phi_t(\xi,x,u,d;\hat{\theta}) \\ \hat{\theta} \end{bmatrix}, \quad t \in \mathbb{T} \setminus \mathbb{T}_y. \vspace{-1.5mm}
\end{equation}
$\textbf{p} \in \mathcal{P}_{\mathbb{X}}$ denotes the prior distribution
of the realizations $\chi_0 \in \mathbb{X}$. $\textbf{p}$ depends on the distribution of $X_0$ 
and the values of $\xi_0$ and $\hat{\theta}_0$ (see just below \eqref{myphit} and \eqref{15a}, respectively).
Let us now study the costs and dynamics presented above. \vspace{-1.5mm}
\begin{theorem}[Regularity of $c_t$, $F_t$, and $G_t$]\label{thm1}
\emph{Let A1 hold. Then,
    $c_t$ \eqref{finalcosts} is l.s.c. and b.b. for every $t \in \mathbb{T}_N$, and
    $F_t$ \eqref{mynewFta} 
    is continuous for every $t \in \mathbb{T}$. In particular, $G_t$ \eqref{myGt} is continuous for every $t \in \mathbb{T} \setminus \mathbb{T}_y$.}
\end{theorem}
\vspace{-10mm}
\hspace{-3mm}\begin{pf*}{Proof.}
The first property holds because $c_t$ is a sum of l.s.c. and b.b. functions, $\hat{c}_t$ or $\hat{c}_t + \lambda \bar{c}_t$. $\lambda \bar{c}_t$ is l.s.c. because $\bar{c}_t$ \eqref{14} is continuous under A1.
$\lambda \bar{c}_t$ is b.b. because
$\lambda \bar{c}_t(\chi) \geq -\lambda b > -\infty$ for every $\chi \in \mathbb{X}$.
For the second property, note that $H_t$ \eqref{Ht} is continuous for every $t \in \mathbb{T}_N$ because it is either a sum of continuous functions or it is constant.
Since $F_t$ is a composition of $\bar{F}_t$ and $H_t$, it suffices to show that $\bar{F}_t = (f_t^\top,\phi_t^\top,g_t^\top)^\top$ is continuous 
in each component. The continuity of $f_t$ \eqref{gensys} and $\phi_t$ \eqref{myphit} follow from A1.
There are two cases for $g_t$ \eqref{15a}.
If $t \in \mathbb{T} \setminus \mathbb{T}_y$, then $g_t(y,x;\hat{\theta}) = \hat{\theta}$, which is continuous. Otherwise, 
$g_t = \bar{g}_t$ \eqref{17} depends on a continuous function $\varphi_\epsilon$, a matrix inverse, and a gradient. The map $(x, \hat{\theta}) \mapsto L_t(x; \hat{\theta})^{-1}$ is continuous because $L_t$ is defined by \eqref{myRt}, $L_t(x; \hat{\theta})$ is positive definite for any $(x,\hat{\theta}) \in \mathbb{S} \times \mathbb{P}$, and
$\frac{\partial h_t}{\partial \theta}$
is continuous 
under A1. The gradient $\text{D}_\theta \varepsilon_t(y,x;\hat{\theta}) \in \mathbb{R}^p$ is given by \vspace{-3mm}
\begin{equation}
   \textstyle \text{D}_\theta \varepsilon_t(y,x;\hat{\theta}) = -2 \left( \frac{\partial h_t(x;\hat{\theta})}{\partial \theta} \right)^\top (y - h_t(x;\hat{\theta}) ). \vspace{-3mm}
\end{equation}
The continuity of $\frac{\partial h_t}{\partial \theta}$ and $h_t$ under A1 imply the continuity of $(y,x,\hat{\theta}) \mapsto \text{D}_\theta \varepsilon_t(y,x;\hat{\theta})$.
Lastly, the continuity of $\varphi_\epsilon$ and $(y, x, \hat{\theta}) \mapsto L_t(x; \hat{\theta})^{-1} \cdot \text{D}_\theta \varepsilon_t(y,x;\hat{\theta})$ imply the continuity of $\bar{g}_t$ \eqref{17}.
\qed
\end{pf*}
\vspace{-7mm}

\section{Representation as a regular POMDP}\label{secIII}\vspace{-3mm}
Here, we show that the system \eqref{augsys} equipped with the information-theoretic costs \eqref{finalcosts} is a nonstationary POMDP with regularity properties under A1. This representation will facilitate policy synthesis in Section \ref{secexistence}.
Next, we formalize the POMDP model of interest. 
\vspace{-1.5mm}
\begin{definition}[Multi-time-scale regular POMDP]\label{pomdpdef}
\emph{A multi-time-scale regular POMDP 
consists of the following: (i) a state space $\mathbb{X}$, compact control space $\mathbb{C}$, 
and measurement space $\mathbb{Y}$, which are nonempty Borel spaces;
(ii) finite discrete-time horizons for the states and measurements, $\mathbb{T}_N$ and $\mathbb{T}_y$, respectively, satisfying the definitions of Section \ref{sysdescription}; 
(iii) stage and terminal cost functions $c_t$ that are l.s.c. and b.b. for every $t \in \mathbb{T}_N$; 
(iv) a distribution $\textbf{p} \in \mathcal{P}_{\mathbb{X}}$ of the initial state;
(v) state transition kernels $q_t$ on $\mathbb{X}$ given $\mathbb{X} \times \mathbb{C}$ that are weakly continuous for every $t \in \{1,2,\dots,N\}$; and 
(vi) observation kernels $s_t$ on $\mathbb{Y}$ given $\mathbb{X}$ that are continuous in total variation for every $t \in \mathbb{T}_y$.}
\end{definition}
\vspace{-1.5mm}
 
Under A1, the system \eqref{augsys} equipped with $c_t$ \eqref{finalcosts} satisfies Parts (i)--(iv) of Definition \ref{pomdpdef}.
We will define state transition and observation kernels and then show that they satisfy Parts (v)--(vi) of Definition \ref{pomdpdef} under A1.
$\chi_t \in \mathbb{X}$, $u_t \in \mathbb{C}$, and $y_t \in \mathbb{Y}$ are realizations of random objects $\mathcal{X}_t$, $U_t$, and $Y_t$, respectively.
For every $t \in \mathbb{T}$, the state transition kernel $q_{t+1}$ provides the conditional distribution of $\mathcal{X}_{t+1}$ given a realization of $(\mathcal{X}_{t},U_{t})$. 
\vspace{-1.5mm}

\begin{definition}[State transition kernels $q_1,\dots, q_N$] \label{qtt}
\emph{For every $t \in \mathbb{T}_y \setminus \{N\}$, $q_{t+1}$ depends on the joint distribution $\nu_t$ of the process and measurement noise $(D_t,W_t)$ as follows: $q_{t+1}(\underline{\mathbb{X}}|\chi,u) \coloneqq$  \vspace{-1mm}
\begin{align*}
   \nu_t(\{(d,w) \in \mathbb{D} \times \mathbb{Y} : F_t(\chi,u,d,w) \in \underline{\mathbb{X}} \}), \;\; t \in \mathbb{T}_y \setminus \{N\}, \vspace{-5mm}
\end{align*}
for every $\underline{\mathbb{X}} \in \mathcal{B}_{\mathbb{X}}$ and $(\chi,u) \in \mathbb{X} \times \mathbb{C}$. Otherwise, if $t \in \mathbb{T} \setminus \mathbb{T}_y$, then there is no measurement and thus no measurement noise. In this case, $q_{t+1}$ depends on the distribution $\mu_t$ of $D_t$ and the function $G_t$ \eqref{myGt}:\vspace{-1.5mm}
\begin{equation*}
  q_{t+1}(\underline{\mathbb{X}}|\chi,u)\coloneqq  \mu_t(\{d \in \mathbb{D} : G_t(\chi,u,d) \in \underline{\mathbb{X}} \}), \;\; t \in \mathbb{T} \setminus \mathbb{T}_y,\vspace{-2mm}
\end{equation*}
for every $\underline{\mathbb{X}} \in \mathcal{B}_{\mathbb{X}}$ and $(\chi,u) \in \mathbb{X} \times \mathbb{C}$.}
\end{definition}
\vspace{-1.5mm}

$q_{t}$ differs from a typical state transition kernel in two ways. Its definition accounts for the absence of the measurement noise at particular time points and how this absence leads to dynamics functions with different dependencies. The observation kernel $s_t$ is simpler because it only pertains to $t \in \mathbb{T}_y$. For every $t \in \mathbb{T}_y$,
$s_t$ provides the conditional distribution of $Y_t$ given a realization of $\mathcal{X}_t$. \vspace{-1.5mm}

\begin{definition}[Observation kernel $s_t$ for $t\in \mathbb{T}_y$] \label{st}
\emph{$s_t$ depends on the distribution $\rho_t$ of $W_t$ as follows:\vspace{-3mm} 
\begin{equation*}
   s_t(\underline{\mathbb{Y}}|\chi) \coloneqq \rho_t(\{w \in \mathbb{Y} : h_t(x; \hat{\theta}) + w \in \underline{\mathbb{Y}} \}), \quad t \in \mathbb{T}_y,\vspace{-3mm} 
\end{equation*}
for every $\underline{\mathbb{Y}} \in \mathcal{B}_{\mathbb{Y}}$ and $\chi = (x^\top,\xi^\top,\hat{\theta}^\top)^\top \in \mathbb{X}$.} 
\end{definition}

In the POMDP literature, the observation kernel commonly depends on $u_t$. We do not allow $s_t$ to depend on $u_t$ to simplify the output sensitivity matrix \eqref{outputsen}. \vspace{-1.5mm}

\begin{lemma}[Regularity of $q_t$ and $s_t$]\label{lemma1}
\emph{Let A1 hold. Then, $q_t$ of Definition \ref{qtt} and $s_t$ of Definition \ref{st} satisfy Part (v) and Part (vi) of Definition \ref{pomdpdef}, respectively.}
\end{lemma}
\vspace{-10mm}
\hspace{-3mm}\begin{pf*}{Proof.}
For every $t \in \mathbb{T}$, the weak continuity of $q_{t+1}$ follows from the continuity of $F_t$ \eqref{mynewFta} under A1 (Theorem \ref{thm1}). 
For every continuous bounded $\psi : \mathbb{X} \rightarrow \mathbb{R}$, $(\chi,u) \in \mathbb{X} \times \mathbb{C}$, and $t \in \mathbb{T}$, we have that $\int_{\mathbb{X}} \psi \; \mathrm{d}q_{t+1}(\cdot|\chi,u) =$\vspace{-2mm}
\begin{align}\label{changeofmeasure}
    \begin{cases}\int_{\mathbb{D} \times \mathbb{Y}}  \psi(F_t(\chi,u,\cdot,\cdot)) \; \mathrm{d}\nu_t, & \text{if }t \in \mathbb{T}_y \setminus \{N\}, \\ \int_{\mathbb{D}}  \psi(G_t(\chi,u,\cdot)) \; \mathrm{d}\mu_t, & \text{if }t \in \mathbb{T} \setminus \mathbb{T}_y,  \end{cases}\vspace{-6mm}
\end{align}
by a change-of-measure theorem \cite[Th. 1.6.12]{ash1972probability}.
Now, let $(\chi_n,u_n)_{n \in \mathbb{N}}$ in $\mathbb{X} \times \mathbb{C}$ converge to $(\chi,u) \in \mathbb{X} \times \mathbb{C}$, and let $\psi : \mathbb{X} \rightarrow \mathbb{R}$ be continuous and bounded. 
It suffices to show that $\int_{\mathbb{X}} \psi \; \mathrm{d}q_{t+1}(\cdot|\chi_n,u_n) \rightarrow \int_{\mathbb{X}} \psi \; \mathrm{d}q_{t+1}(\cdot|\chi,u)$ in $\mathbb{R}$ as $n \rightarrow +\infty$ (Rmk. \ref{weakconvremark}). First, consider $t \in \mathbb{T}_y \setminus \{N\}$. Define the function $\psi_{t,n} : \mathbb{D} \times \mathbb{Y} \rightarrow \mathbb{R}$ by $\psi_{t,n}  \coloneqq \psi(F_t(\chi_n,u_n,\cdot,\cdot))$
for every $n \in \mathbb{N}$, and define $\psi_t : \mathbb{D} \times \mathbb{Y} \rightarrow \mathbb{R}$ by $\psi_t  \coloneqq \psi(F_t(\chi,u,\cdot,\cdot))$.
We have that $\int_{\mathbb{D} \times \mathbb{Y}} \psi_{t,n} \; \mathrm{d}\nu_t = 
\int_{\mathbb{X}} \psi \; \mathrm{d}q_{t+1}(\cdot|\chi_n,u_n)$ for every $n \in \mathbb{N}$ and $\int_{\mathbb{D} \times \mathbb{Y}} \psi_{t} \; \mathrm{d}\nu_t = \int_{\mathbb{X}} \psi \; \mathrm{d}q_{t+1}(\cdot|\chi,u)$
by the first case of \eqref{changeofmeasure}. 
Since $(\chi_n,u_n) \rightarrow (\chi,u)$ and $\psi \circ F_t$ is continuous, $\psi_{t,n} \rightarrow \psi_t$ 
pointwise as $n \rightarrow +\infty$. 
Also, $\psi$ is bounded, and $\psi_t$ and $\psi_{t,n}$ for every $n \in \mathbb{N}$ are 
continuous and hence measurable.
Thus, we use the Dominated Convergence Theorem 
to conclude that
$\int_{\mathbb{D} \times \mathbb{Y}} \psi_{t,n} \; \mathrm{d}\nu_t \rightarrow \int_{\mathbb{D} \times \mathbb{Y}} \psi_t \; \mathrm{d}\nu_t$ in $\mathbb{R}$ as $n \rightarrow +\infty$
\cite[p. 49]{ash1972probability}.
The analysis of $q_{t+1}$ for any $t \in \mathbb{T}\setminus \mathbb{T}_y$ is similar, and so, we omit it.
For every $t \in \mathbb{T}_y$, the continuity of $s_t$ in total variation is due to $h_t$ being continuous, the measurement noise being additive, $\rho_t$ admitting a continuous nonnegative density,
and Scheff\'{e}'s Lemma \cite[Sec. 5.10]{williams1991probability}. The reader may see \cite[Sec. 2.2 (ii)]{kara2019weak} or \cite[Remark 5]{chapman2022tac} for more details. 
\qed
\end{pf*}
\vspace{-5.5mm}

Hence, the system \eqref{augsys} with the costs \eqref{finalcosts}, state transition kernels of Definition \ref{qtt}, and observation kernels of Definition \ref{st} is a POMDP that satisfies Definition \ref{pomdpdef} under A1. \vspace{-2.5mm}

\section{Policy Synthesis}\label{secexistence} \vspace{-3mm}
A POMDP can be reduced to a fully observable \emph{belief-space} MDP, whose state space is the space of posterior distributions of the unobservable state \cite{rhenius1974incomplete} \cite{yushkevich1976reduction}. Moreover, if there exists an optimal policy for the belief-space MDP, then an optimal policy for the POMDP can be constructed \cite[Prop. 10.3]{bertsekas2004stochastic} 
\cite[pp. 89--90]{hernandez1989adaptive}.
Thus, the purpose of this section is to show the existence of an optimal policy under A1 for a belief-space MDP corresponding to the POMDP of the previous section. (Verifying the equivalence between the POMDP and belief-space MDP solutions is out of scope of the current brief paper; related proofs can be found in \cite[Prop. 10.4, Prop. 10.5]{bertsekas2004stochastic}, for example.)\vspace{-2mm}
 
First, we formalize our belief-space MDP. We define its state and trajectory spaces (Def. \ref{def4}), random states and controls (Def. \ref{def5}), random cost variable (Def. \ref{myC}), Markov policy class (Def. \ref{defpi}), state transition kernels (Def. \ref{def7}, Def. \ref{def8}), and initial distribution (Def. \ref{initq0}). Then, we provide a policy existence result (Theorem \ref{thmpolicy}).
The definitions are required to state the theorem. While the definitions resemble the standard ones for POMDPs \cite[Ch. 10]{bertsekas2004stochastic} \cite{feinberg2016partially} \cite[Ch. 4]{hernandez1989adaptive},
we must also circumvent the issue of measurements not necessarily being available always. This issue requires us to provide different definitions for the state transition kernels (Def. \ref{def8}) and for the initial distribution (Def. \ref{initq0}).
\vspace{-2mm}

\begin{definition}[$\mathbb{Z}, \Omega$]\label{def4}
\emph{The state space of the belief-space MDP is $\mathbb{Z} \coloneqq \mathcal{P}_{\mathbb{X}}$. A trajectory $\omega \in \Omega \coloneqq (\mathbb{Z} \times \mathbb{C})^N \times \mathbb{Z}$ takes the form $\omega = (z_0,u_0,\dots,z_{N-1},u_{N-1},z_N)$, where the coordinates of $\omega$ are related casually.
}
\end{definition}\vspace{-1.5mm}
\begin{definition}[$Z_t$, $\tilde{U}_t$]\label{def5}
\emph{We define $Z_t : \Omega \rightarrow \mathbb{Z}$ for every $t \in \mathbb{T}_N$ and $\tilde{U}_t : \Omega \rightarrow  \mathbb{C}$ for every $t \in \mathbb{T}$ to be projections, i.e., $Z_t(\omega) \coloneqq z_t$ and $\tilde{U}_t(\omega) \coloneqq u_t$ for every $\omega \in \Omega$ of the form in Definition \ref{def4}.}
\end{definition}\vspace{-1.5mm}

\begin{definition}[$C$]\label{myC}
\emph{Under A1, we define $C : \Omega \rightarrow \mathbb{R}^*$ by
    $C(\omega) \coloneqq \tilde{c}_N(Z_N(\omega)) + \sum_{t = 0}^{N-1} \tilde{c}_t(Z_t(\omega),\tilde{U}_t(\omega))$,
where $\tilde{c}_t$ is related to $c_t$ \eqref{finalcosts} as follows: \vspace{-1mm}
\begin{subequations}\label{tildect}
\begin{align}
    \tilde{c}_t(z_t,u_t) & \coloneqq \textstyle \int_{\mathbb{X}} c_t(\chi_t,u_t) \; z_t(\mathrm{d}\chi_t), \quad t \in \mathbb{T}, \label{16a} \\
    \tilde{c}_N(z_N) & \coloneqq \textstyle \int_{\mathbb{X}} c_N(\chi_N) \; z_N(\mathrm{d}\chi_N), \label{16b}\vspace{-2mm}
\end{align}
\end{subequations}
with $\tilde{c}_t : \mathbb{Z} \times \mathbb{C} \rightarrow \mathbb{R}^*$ for every $t \in \mathbb{T}$ and $\tilde{c}_N : \mathbb{Z} \rightarrow \mathbb{R}^*$.}
\end{definition}\vspace{-1.5mm}

\begin{definition}[$\Pi$, $\mathbb{C}_t$]\label{defpi}
\emph{We define $\mathbb{C}_t \coloneqq \mathbb{C}$ if $t \in \mathbb{T}_u$ and $\mathbb{C}_t \coloneqq\{\mathbf{u}\}$ if $t \in \mathbb{T} \setminus \mathbb{T}_u$. A control policy $\pi = (\pi_0,\pi_1,\dots,\pi_{N-1}) \in \Pi$ is a finite sequence 
of measurable stochastic kernels on $\mathbb{C}$ given $\mathbb{Z}$ such that $\pi_t(\mathbb{C}_t|z_t) = 1$ for every $z_t \in \mathbb{Z}$ and $t \in \mathbb{T}$. $\Pi$ is the set of all policies.}
\end{definition}\vspace{-1.5mm}

In Definitions \ref{def7} and \ref{def8}, we specify the state transition kernels $\tilde{q}_1,\tilde{q}_2,\dots,\tilde{q}_N$ of the belief-space MDP.
For every $t \in \mathbb{T}$, $\tilde{q}_{t+1}(\cdot|z_t,u_t) \in \mathcal{P}_{\mathbb{Z}}$ is the conditional distribution of $Z_{t+1}$ 
given a realization $(z_t,u_t) \in \mathbb{Z} \times \mathbb{C}$ of $(Z_t,\tilde{U}_t)$. Definition \ref{def7} applies when a measurement is present, whereas Definition \ref{def8} applies when a measurement is absent. In each case, some technical lead-up is required before writing the expression for $\tilde{q}_{t+1}$.
\vspace{-1.5mm}

\begin{definition}[$\tilde{q}_{t+1}$ \textnormal{when a measurement is present}]\label{def7}
\emph{If $t+1 \in \mathbb{T}_y\setminus\{0\}$, then $\tilde{q}_{t+1}$ is constructed using a joint conditional distribution $r_{t+1}(\cdot,\cdot|z_t,u_t)$ of $(Y_{t+1},\mathcal{X}_{t+1})$. $r_{t+1}$ is a measurable stochastic kernel on $\mathbb{Y} \times \mathbb{X}$ given $\mathbb{Z} \times \mathbb{C}$
such that $r_{t+1}(\underline{\mathbb{Y}}\times \underline{\mathbb{X}}|z_t,u_t) \coloneqq$ \vspace{-1.5mm}
\begin{equation}\label{rt}
\begin{aligned}
 \textstyle   \int_{\mathbb{X}} \int_{\underline{\mathbb{X}}} s_{t+1}(\underline{\mathbb{Y}}|\chi_{t+1}) \; q_{t+1}(\mathrm{d}\chi_{t+1}|\chi_t,u_t) \; z_t(\mathrm{d}\chi_t)
    \end{aligned}\vspace{-1.5mm}
\end{equation} 
for every $\underline{\mathbb{Y}} \in \mathcal{B}_{\mathbb{Y}}$, $\underline{\mathbb{X}} \in \mathcal{B}_{\mathbb{X}}$, and $(z_t,u_t) \in \mathbb{Z} \times \mathbb{C}$.
Denote the marginal of $r_{t+1}$ on $\mathbb{Y}$ by $r_{t+1}'$, i.e., $r_{t+1}'(\underline{\mathbb{Y}}|z_t,u_t) \coloneqq r_{t+1}(\underline{\mathbb{Y}}\times\mathbb{X}|z_t,u_t)$ for every $\underline{\mathbb{Y}} \in \mathcal{B}_{\mathbb{Y}}$ and $(z_t,u_t) \in \mathbb{Z} \times \mathbb{C}$. 
$r_{t+1}$
enjoys a decomposition in terms of this marginal as follows: for every $\underline{\mathbb{Y}} \in \mathcal{B}_{\mathbb{Y}}$, $\underline{\mathbb{X}} \in \mathcal{B}_{\mathbb{X}}$, and $(z_t,u_t) \in \mathbb{Z} \times \mathbb{C}$, \vspace{-2mm} 
\begin{equation*}
 \textstyle    r_{t+1}(\underline{\mathbb{Y}}\times \underline{\mathbb{X}}|z_t,u_t)  = \int_{\underline{\mathbb{Y}}} \Phi_{t+1}(\underline{\mathbb{X}}|z_t,u_t,y)\; r_{t+1}'(\mathrm{d}y|z_t,u_t), \vspace{-2mm}
\end{equation*}
where $\Phi_{t+1}$ is a measurable stochastic kernel on $\mathbb{X}$ given $\mathbb{Z} \times \mathbb{C} \times \mathbb{Y}$ \cite[Cor. 7.27.1]{bertsekas2004stochastic}. 
Then, we define $\tilde{q}_{t+1}(\underline{\mathbb{Z}}|z_t,u_t) \coloneqq $\vspace{-1.5mm}
\begin{equation}\label{27}
   r_{t+1}'(\{y_{t+1} \in \mathbb{Y} : \Phi_{t+1}(\cdot|z_t,u_t,y_{t+1}) \in \underline{\mathbb{Z}} \}|z_t,u_t) \vspace{-1.5mm} 
\end{equation}
for every $\underline{\mathbb{Z}} \in \mathcal{B}_{\mathbb{Z}}$ and $(z_t,u_t) \in \mathbb{Z} \times \mathbb{C}$.}
\end{definition}

\begin{definition}[$\tilde{q}_{t+1}$ \textnormal{when a measurement is absent}]\label{def8}
\emph{If $t+1 \in \{1,2,\dots,N\}\setminus \mathbb{T}_y$, 
then $\tilde{q}_{t+1}$ concentrates the realizations of $Z_{t+1}$ at a prior conditional distribution $\eta_{t+1}(\cdot|z_t,u_t) \in \mathbb{Z}$ of $\mathcal{X}_{t+1}$; i.e., we define $\tilde{q}_{t+1}$ by \vspace{-1.5mm}
\begin{equation}\label{29}
    \tilde{q}_{t+1}(\underline{\mathbb{Z}}|z_t,u_t) \coloneqq \delta_{\eta_{t+1}(\cdot|z_t,u_t)}(\underline{\mathbb{Z}}) \vspace{-1.5mm}
\end{equation}
for every $\underline{\mathbb{Z}} \in \mathcal{B}_{\mathbb{Z}}$ and $(z_t,u_t) \in \mathbb{Z} \times \mathbb{C}$,
where
$\eta_{t+1}$ is
a measurable stochastic kernel on $\mathbb{X}$ given $\mathbb{Z} \times \mathbb{C}$, which is defined by \cite[p. 261, Eq. (50)]{bertsekas2004stochastic} \vspace{-1.5mm}
\begin{equation}\label{28}
 \textstyle   \eta_{t+1}(\underline{\mathbb{X}}|z_t,u_t) \coloneqq 
      \int_{\mathbb{X}}  \; q_{t+1}(\underline{\mathbb{X}}|\chi_t,u_t) \; z_t(\mathrm{d}\chi_t) \vspace{-1.5mm}
\end{equation}
for every $\underline{\mathbb{X}} \in \mathcal{B}_{\mathbb{X}}$ and $(z_t,u_t) \in \mathbb{Z} \times \mathbb{C}$.
}\vspace{-1.5mm}
\end{definition}
For $t+1 \in \{1,2,\dots,N\}$, the expression for $\tilde{q}_{t+1}$ is in \eqref{27} or \eqref{29}. Lastly, we specify the initial kernel $\tilde{q}_0$, whose construction also depends on the availability of a measurement. $\tilde{q}_0(\cdot|\textbf{p}) \in \mathcal{P}_{\mathbb{Z}}$ is the distribution of $Z_0$ when $\textbf{p} \in \mathbb{Z}$ is the distribution of $\mathcal{X}_0$.  \vspace{-1mm}
\begin{definition}[$\tilde{q}_0$]\label{initq0}
\emph{If $0 \in \mathbb{T}_y$, then $\tilde{q}_0$ depends on a joint distribution $r_0(\cdot,\cdot|\textbf{p})$ of $(Y_0,\mathcal{X}_0)$, where $r_0$ is defined by \vspace{-2mm}
\begin{align}\label{30}
    r_0(\underline{\mathbb{Y}} \times \underline{\mathbb{X}}|\textbf{p}) & \coloneqq \textstyle \int_{\underline{\mathbb{X}}} s_{0}(\underline{\mathbb{Y}}|\chi_{0}) \; \textbf{p}(\mathrm{d}\chi_0) \\
     & \;= \textstyle \int_{\underline{\mathbb{Y}}} \Phi_0(\underline{\mathbb{X}}|\textbf{p},y_0)\; r_0'(\mathrm{d}y_0|\textbf{p})  \vspace{-3mm}
\end{align}
for every $\underline{\mathbb{Y}} \in \mathcal{B}_{\mathbb{Y}}$, $\underline{\mathbb{X}} \in \mathcal{B}_{\mathbb{X}}$, and $\textbf{p} \in \mathbb{Z}$, $r_0'$ is defined by
$r_0'(\underline{\mathbb{Y}}|\textbf{p}) \coloneqq r_0(\underline{\mathbb{Y}}\times\mathbb{X}|\textbf{p})$ for every $\underline{\mathbb{Y}} \in \mathcal{B}_{\mathbb{Y}}$ and $\textbf{p} \in \mathbb{Z}$, and $\Phi_0$ is a measurable stochastic kernel on $\mathbb{X}$ given $\mathbb{Z} \times \mathbb{Y}$ \cite[Cor. 7.27.1]{bertsekas2004stochastic}. 
However, if $0 \notin \mathbb{T}_y$, then $\tilde{q}_0(\cdot|\textbf{p})$ concentrates the realizations of $Z_0$ at $\textbf{p}$. All together, we define $\tilde{q}_0(\underline{\mathbb{Z}}|\textbf{p}) \coloneqq$\vspace{-2mm}
\begin{equation}
    \begin{cases} r_0'(\{y_0 \in \mathbb{Y} : \Phi_0(\cdot|\textbf{p},y_0) \in \underline{\mathbb{Z}} \} |\textbf{p}), & \text{if }0 \in \mathbb{T}_y, \\ 
    \delta_{\textbf{p}}(\underline{\mathbb{Z}}), &  \text{if }0 \notin \mathbb{T}_y,\end{cases}\vspace{-2mm}
\end{equation}
for every $\underline{\mathbb{Z}} \in \mathcal{B}_{\mathbb{Z}}$ and $\textbf{p} \in \mathbb{Z}$.
}
\end{definition}

Equipped with the above definitions, now we can specify a probability measure, an expected cost, and the optimal control problem of interest. By the Ionescu-Tulcea Theorem, 
for every $\pi \in \Pi$ and $\textbf{p} \in \mathbb{Z}$, there is a unique probability measure $P_{\tilde{q}_0(\cdot|\textbf{p})}^\pi \in \mathcal{P}_{\Omega}$ that depends on  
$\pi$, $\tilde{q}_0(\cdot|\textbf{p})$, 
and 
$\tilde{q}_{t+1}$ for all $t \in \mathbb{T}$.
If $G : \Omega \rightarrow \mathbb{R}^*$ is measurable and b.b., then the expectation of $G$ with respect to $P_{\tilde{q}_0(\cdot|\textbf{p})}^\pi$ is defined by $E_{\tilde{q}_0(\cdot|\textbf{p})}^\pi(G) \coloneqq \int_{\Omega} G \; \mathrm{d}P_{\tilde{q}_0(\cdot|\textbf{p})}^\pi$. 
The next result guarantees that an optimal policy $\pi^* \in \Pi$ for the belief-space MDP exists under A1, i.e.,\vspace{-3mm}
\begin{equation}\label{opt}
   E_{\tilde{q}_0(\cdot|\textbf{p})}^{\pi^*}(C) = J^*(\textbf{p}) \coloneqq \inf_{\pi \in \Pi} E_{\tilde{q}_0(\cdot|\textbf{p})}^\pi(C), \quad \textbf{p} \in \mathbb{Z}.\vspace{-1mm}
\end{equation}

\begin{theorem}[Policy synthesis]\label{thmpolicy}
\emph{Let A1 hold. Define $J_N \coloneqq \tilde{c}_N$ \eqref{16b}, and for $t = N-1,\dots,1,0$, define $J_t : \mathbb{Z} \rightarrow \mathbb{R}^*$ recursively backwards in time by\vspace{-1.5mm}
\begin{equation}\label{Jt}
J_t(z_t) \coloneqq \underset{u_t \in \mathbb{C}_t}{\inf} V_t(z_t,u_t) ,
    \vspace{-1.5mm}
\end{equation}
where $\mathbb{C}_t$ is from Def. \ref{defpi}, $V_t \coloneqq \tilde{c}_t + V_{t}'$, $\tilde{c}_t$ is defined by \eqref{16a}, and $V_{t}': \mathbb{Z} \times \mathbb{C} \rightarrow \mathbb{R}^*$ depends on $J_{t+1}$ as follows: \vspace{-1mm}
\begin{align}
    V_{t}'(z_t,u_t) & \coloneqq \textstyle \int_{\mathbb{Z}} J_{t+1}(z_{t+1}) \; \tilde{q}_{t+1}(\mathrm{d}z_{t+1}|z_t,u_t). \label{Vtprime} \vspace{-5mm}
\end{align}
Then, for every $t \in \mathbb{T}_N$, $J_t$ is l.s.c. and b.b.
Also, for every $t \in \mathbb{T}$, there is a measurable function $\kappa_t^* : \mathbb{Z} \rightarrow \mathbb{C}$ such that $\kappa_t^*(z_t) \in \mathbb{C}_t$ and $J_t(z_t) = V_t(z_t,\kappa_t^*(z_t))$ for every $z_t \in \mathbb{Z}$. In particular, for every $t \in \mathbb{T}\setminus \mathbb{T}_u$, we choose $\kappa_t^*(z_t) \coloneqq \textbf{u}$ for every $z_t \in \mathbb{Z}$. Lastly, $\pi^* \coloneqq (\delta_{\kappa_0^*}, \delta_{\kappa_1^*}, \dots,\delta_{\kappa_{N-1}^*}) \in \Pi$ (Rmk. \ref{remark4}) is optimal for the belief-space MDP, i.e., $\pi^*$ satisfies \eqref{opt}.}
\end{theorem}
\vspace{-6mm}
\begin{pf*}{Proof.}
We will prove the first two statements by induction. The arguments have nuances due to the two cases for $\tilde{q}_{t+1}$ \eqref{27} \eqref{29}. The last statement holds by a typical dynamic programming argument, e.g., see \cite[Ch. 8]{bertsekas2004stochastic} or \cite[Ch. 3]{hernandez2012discrete}. 
\vspace{-2mm}

Under A1, for every $t \in \mathbb{T}_N$, $\tilde{c}_t$ \eqref{tildect} is l.s.c. and b.b. because $c_t$ \eqref{finalcosts} enjoys these properties (Theorem \ref{thm1}) and due to the Generalized Fatou's Lemma \cite[Lemma 6.1]{feinberg2016partially}. 
Now, $J_N = \tilde{c}_N$ is l.s.c. and b.b. Assume the induction hypothesis: for some $t \in \mathbb{T}$,
$J_{t+1}$ is l.s.c. and b.b. If $V_{t}'$ \eqref{Vtprime} 
is l.s.c. and b.b., then $J_t$ \eqref{Jt} is l.s.c. and b.b., and there is a measurable function $\kappa_t^*$ that satisfies the properties specified in the statement of Theorem \ref{thmpolicy} (A1, Rmk. \ref{measselection}).
If $\tilde{q}_{t+1}$ is weakly continuous, then $V_{t}'$ \eqref{Vtprime} is l.s.c. and b.b. (induction hypothesis, Rmk. \ref{remark3}).  \vspace{-2mm}

Hence, 
it suffices to show that $\tilde{q}_{t+1}$ is weakly continuous. 
First, consider $t+1 \in \mathbb{T}_y \setminus \{0\}$. Then, $q_{t+1}$ (Def. \ref{qtt}) is weakly continuous, and $s_{t+1}$ (Def. \ref{st}) is continuous in total variation under A1 (Lemma \ref{lemma1}).
These two continuity properties imply that $\tilde{q}_{t+1}$ \eqref{27} is weakly continuous directly from \cite[Th. 1]{kara2019weak}, which first appeared in \cite{feinberg2016partially}.
Otherwise, if $t+1 \in \{1,2,\dots,N\} \setminus \mathbb{T}_y$, then $\tilde{q}_{t+1}$ is defined by \eqref{29}. We must show that
$(z,u) \mapsto \tilde{q}_{t+1}(\cdot|z,u) = \delta_{\eta_{t+1}(\cdot|z,u)}$ is weakly continuous. 
We define $\ell : \mathbb{Z} \rightarrow \mathcal{P}_{\mathbb{Z}}$ by $\ell(z) \coloneqq \delta_z$ and $v : \mathbb{Z} \times \mathbb{C} \rightarrow \mathbb{Z}$ by $v(z,u) \coloneqq \eta_{t+1}(\cdot|z,u)$, and thus,
    $\tilde{q}_{t+1}(\cdot|z,u) = \ell(v(z,u))$.
Also, $\ell$ is weakly continuous (Rmk. \ref{remark3}). Hence, 
it suffices to show that $v$ is weakly continuous. 
More specifically,
it suffices to show that for any continuous bounded function $\psi : \mathbb{X} \rightarrow \mathbb{R}$, the function $\psi' : \mathbb{Z} \times \mathbb{C} \rightarrow \mathbb{R}$ defined by
   $ \psi'(z,u)  \coloneqq \int_{\mathbb{X}} \psi \; \mathrm{d}\eta_{t+1}(\cdot|z,u)$
is continuous (Rmk. \ref{weakconvremark}). By the definition of $\eta_{t+1}$ \eqref{28}, we have that \vspace{-2mm}
\begin{equation}\label{psiprime}
    \psi'(z,u) = \textstyle \int_{\mathbb{X}} \int_{\mathbb{X}} \psi(a) \;q_{t+1}(\mathrm{d}a|\chi,u) \; z(\mathrm{d}\chi). \vspace{-2mm}
\end{equation}
We will re-express $\psi'$ \eqref{psiprime} to apply the weak continuity of products of measures in $\mathcal{P}_{\mathbb{X}}$ and $\mathcal{P}_{\mathbb{C}}$.
Define $\Psi : \mathbb{X} \times \mathbb{C} \times \mathbb{X} \rightarrow \mathbb{R}$ by $\Psi(\chi, u, \cdot) \coloneqq \psi$ for all $(\chi, u) \in \mathbb{X} \times \mathbb{C}$.
Next, define $\vartheta : \mathbb{X} \times \mathbb{C} \rightarrow \mathbb{R}$ by\vspace{-2mm}
\begin{equation}\label{myinnerfunc}
    \vartheta(\chi,u) \coloneqq \textstyle \int_{\mathbb{X}} \Psi(\chi, u, a) \;q_{t+1}(\mathrm{d}a|\chi,u), \vspace{-2mm}
\end{equation}
which is continuous and bounded (Lemma \ref{lemma1}, Rmk. \ref{remark3}). 
Now, $\psi'$ \eqref{psiprime} can be expressed in terms of $\vartheta$ \eqref{myinnerfunc}: \vspace{-.5mm}
\begin{align}
    \psi'(z,u)
    & = \textstyle \int_{\mathbb{X}} \int_{\mathbb{C}} \vartheta(\chi,\bar{u}) \; \delta_{u}(\mathrm{d}\bar{u})\; z(\mathrm{d}\chi) \\ & = \textstyle \int_{\mathbb{X} \times \mathbb{C}} \vartheta \; \mathrm{d}(z  \delta_{u}), \label{43} \vspace{-4mm}
\end{align}
where $z  \delta_{u} \in \mathcal{P}_{\mathbb{X} \times \mathbb{C}}$ is the product of $z \in \mathcal{P}_{\mathbb{X}}$ and $\delta_{u} \in \mathcal{P}_{\mathbb{C}}$. 
Let $z_n \rightarrow z$ weakly in $\mathbb{Z}$, and let $u_n \rightarrow u$ in $\mathbb{C}$.
The latter 
implies that $\delta_{u_n} \rightarrow \delta_{u}$ weakly in $\mathcal{P}_{\mathbb{C}}$ (Rmk. \ref{remark3}).
The weak convergence of  $z_n \rightarrow z$ in
$\mathcal{P}_{\mathbb{X}}$ 
and $\delta_{u_n} \rightarrow \delta_{u}$ in $\mathcal{P}_{\mathbb{C}}$
implies the weak convergence of $z_n \delta_{u_n}  \rightarrow   z \delta_{u}$ in $\mathcal{P}_{\mathbb{X} \times \mathbb{C}}$ (Rmk. \ref{remark3}).
The weak convergence of these products is equivalent to the convergence of
$\int_{\mathbb{X} \times \mathbb{C}} h \; \mathrm{d}(z_n \delta_{u_n})\rightarrow \int_{\mathbb{X} \times \mathbb{C}} h \; \mathrm{d}(z \delta_{u})$ in $\mathbb{R}$ for every continuous bounded function $h: \mathbb{X} \times \mathbb{C} \rightarrow \mathbb{R}$ (Rmk. \ref{weakconvremark}).
By considering $h = \vartheta$ \eqref{myinnerfunc}, we find that $\psi'(z_n,u_n) \rightarrow \psi'(z,u)$ in $\mathbb{R}$ in view of \eqref{43}. \qed
\end{pf*}

\vspace{-6.5mm}
This concludes the theoretical portion of the brief paper. \vspace{-7mm}

\section{Conceptual example}\vspace{-4mm}\label{example}
In this section, we provide a conceptual leukemia treatment model and explain why it satisfies A1.
We consider the setting of adjusting the dose of an oral chemotherapy called 6-Mercaptopurine (6-MP)
during a treatment cycle for a leukemia patient. Currently, the dose is adjusted according to the current neutrophil concentration (a neutrophil is a type of white blood cell). The concentration is measured at least once per cycle,
whose typical length is 21 days. If the concentration
is outside a desired range, then the dose of 6-MP is
modified (increased or decreased by 20\%) for the next
cycle. Nominally, the drug is taken at the prescribed
dose for the first 14 days, and then no drug is taken for
the remaining 7 days. \vspace{-1.5mm} 

Continuous-time models for the kinetics of 6-MP
\cite{jayachandran2015model} \cite{jayachandran2014optimal}
and their effect on white blood cells \cite{jayachandran2014optimal} \cite{karppinen2019prediction}
have been developed recently. 
The models are deterministic in \cite{jayachandran2015model} \cite{jayachandran2014optimal}. Three states represent the amounts of 6-MP-derived substances in \cite{jayachandran2015model}. 
Five states represent the amounts of white blood cells in different stages of maturity under the influence of 6-MP in \cite{jayachandran2014optimal}.
A stochastic model with two states that neglects the delayed effect of 6-MP on white blood cells 
has been proposed in \cite{karppinen2019prediction}.
These models represent one or both of the following biochemical processes: 1) how 6-MP is broken down into a substance called 6-thioguanine nucleotide (6-TGN);
2) how 6-TGN
influences the proliferation and maturation of white blood cells in the bone marrow. 
\vspace{-2mm}

\begin{figure}[t]
    \centering
    \includegraphics[width = \columnwidth]{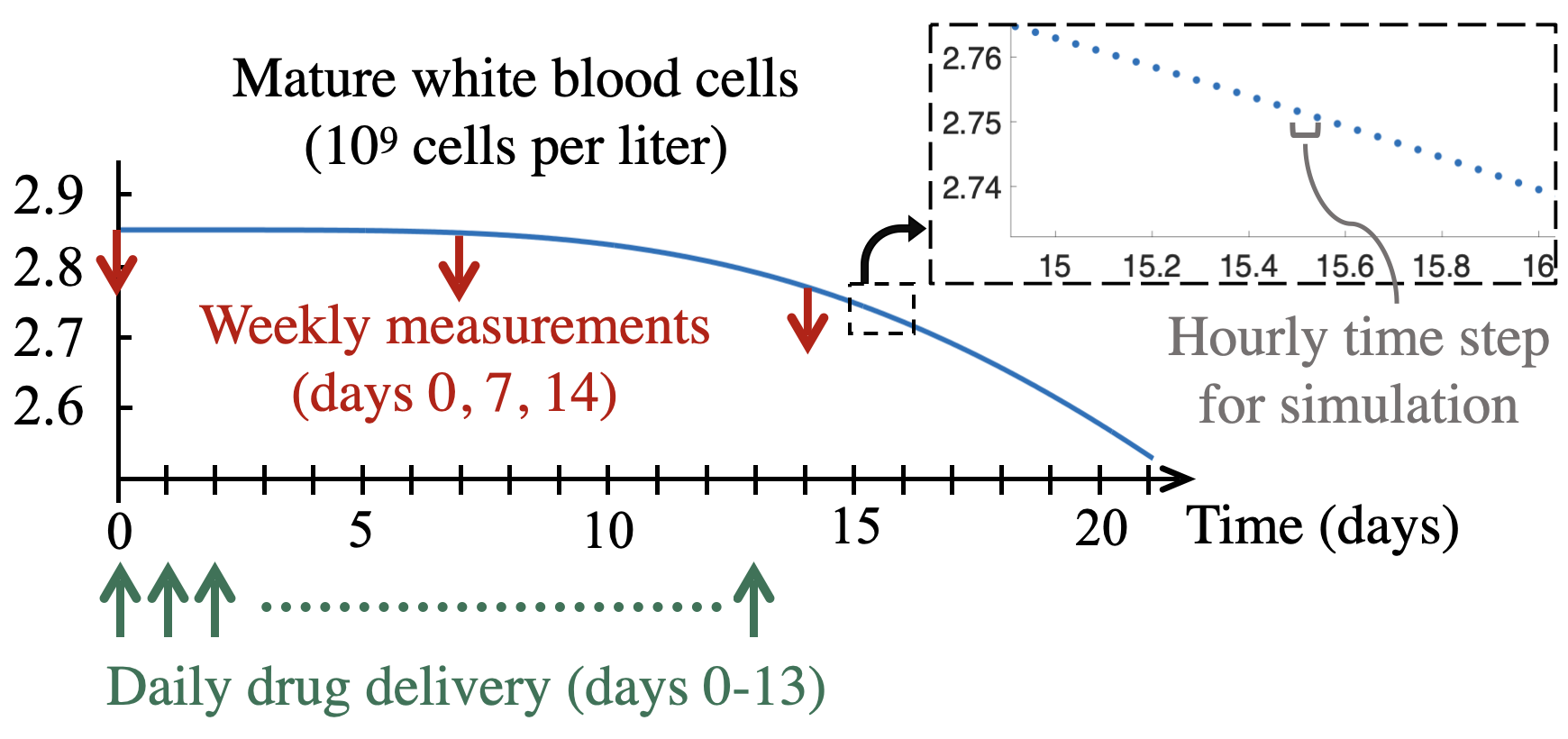}\vspace{-2.5mm}
    \caption{{An illustration of the time scales in the conceptual leukemia treatment example. The close-up illustrates $\mathbb{T}$, where the time steps are hourly. For the illustration, we have simulated the mature white blood cells using a nominal model (to be presented in Eq. \eqref{fbarexample}) and the dosing regimen from \cite{deangelo2015long}. In the simulation, we have initialized the states to be the equilibrium values of the dynamics when the drug input is zero; parameter values are from \cite{jayachandran2014optimal, jayachandran2015model}.}}
    \label{fig:timeScales}
\end{figure}

Building on the above models, we consider a discrete-time stochastic model for the two biochemical processes.
The length of $\mathbb{T}$ is the length of a typical cycle (21 days).
$\mathbb{T}_u$ equals the time points corresponding to the first 14 days. If weekly measurements are planned, then $\mathbb{T}_y$ equals the time points corresponding to days 0, 7, and 14. Figure \ref{fig:timeScales} illustrates the different time scales. The control space $\mathbb{C} = [0,\bar{u}] \subset \mathbb{R}$ is compact (A1 (c)), where $\bar{u} \in \mathbb{R}_+$ is the maximum permissible dose rate. 
We may choose $\bar{u}$ (mg/day) to be twice the nominal dose rate of $50$ (mg/($\text{m}^2 \cdot \text{day}$)), i.e., $\bar{u} = 2 \cdot 50 \cdot \text{BSA}$, where $\text{BSA}$ ($\text{m}^2$) is the patient's body surface area. The default value is $\textbf{u} = 0$, which refers to no drug being taken.
We assume Gaussian process and measurement noise due to the general role of noise in biochemical systems \cite{rao2002control} \cite{eling2019challenges} and lack of domain-specific knowledge about the noise distributions. This is one way to satisfy the measurement noise condition (A1 (e)).
    In our model, $x_1$, $x_2$, and $x_3$ are the three states representing the amounts of 6-MP-derived substances from \cite{jayachandran2015model}, while $x_4$ through $x_8$ are the five states representing the amounts of white blood cells in different stages of maturity under the influence of 6-MP from \cite{jayachandran2014optimal} (Table \ref{statestable}).
The parameters include
various rates,
drug-effect quantities, and the ratio between neutrophils and mature white blood cells (Table \ref{statestable}). The measurements are neutrophils and mature white blood cells.
The spaces of interest are
    $\mathbb{S} = \mathbb{R}_+^8$, $\mathbb{P} = \mathbb{R}_+^{13}$, $\mathbb{C} = [0,\bar{u}]$, $\mathbb{D} = \mathbb{R}^8$, and $\mathbb{Y} = \mathbb{R}^2$. 
\begin{table}[h]
\setlength{\tabcolsep}{3pt}
\begin{tabular}{ |p{40pt}|p{185pt}| } 
 \hline
{ Symbol} & { Description} \\
\hline 
$x_1$ & Amount of 6-MP in the gut  \\
$x_2$ & Amount of 6-MP in the plasma  \\
$x_3$ & Concentration of 6-TGN in red blood cells \\
$x_4$ & Concentration of proliferating white blood cells \\
$x_5$, $x_6$, $x_7$ & Concentrations of maturing white blood cells \\
$x_8$ & Concentration of mature white blood cells\\
$\theta_1$ & Conversion rate (6-MP to 6-TGN)   \\
$\theta_2$ &  Michaelis-Menten constant  \\
$\theta_3$ &  Maximum proliferation rate \\
$\theta_4$ & Steepness parameter  \\
$\theta_5$ & Feedback parameter  \\
$\theta_6$ & Maximum drug effect on mature white blood cells  \\
$\theta_7$ & Saturation constant for drug effect \\
$\theta_8,\dots,\theta_{12}$ & Rate of absorption, elimination, or transition \\
$\theta_{13}$ & Ratio between neutrophils and mature white blood cells \\
 \hline 
\end{tabular}
\caption{}
\label{statestable} 
\end{table}\vspace{-4mm}

Letting $\Delta \in \mathbb{R}_+$ be the duration of $[t,t+1)$ for every $t \in \mathbb{T}$ and $\hat{\varphi}_\epsilon : \mathbb{R}^8 \rightarrow \mathbb{S}$ be a continuously differentiable approximation for $\varphi_\epsilon : \mathbb{R}^8 \rightarrow \mathbb{S}$ (Sec. \ref{notationsection}),
\footnote{One choice is $\max\{x_i,\epsilon\} \approx \hat{\varphi}_{\epsilon,\beta}(x_i) \coloneqq \beta^{-1} \log( \exp(\beta x_i) + \exp(\beta \epsilon) )$ with $\beta \geq 1$, which satisfies
   $ \max\{x_i, \epsilon\} \leq \hat{\varphi}_{\epsilon,\beta}(x_i) \leq \max\{x_i, \epsilon\} + \beta^{-1}\log(2)$.}
we consider
\vspace{-1.5mm}
\begin{align}
    f_t(x,u,d;\theta) & = \hat{\varphi}_\epsilon( x + \Delta \cdot \bar{f}(x,u; \theta) + d),\label{ftexample}\\
    h_t(x;\theta) & = C(\theta) x, \label{htexample}\\
    \bar{f}(x,u; \theta) & \coloneqq A(\theta)  x + B  u + \hat{f}(x; \theta), \vspace{-2.5mm}\label{fbarexample}
\end{align}
where $A(\theta) \in \mathbb{R}^{8 \times 8}$ and $C(\theta) \in \mathbb{R}^{2 \times 8}$ depend linearly on $\theta$, $B \in \mathbb{R}^8$ is constant, and $\hat{f}$ is nonlinear. 
The definition of $\bar{f}$ \eqref{fbarexample} comes from \cite{jayachandran2015model} and \cite{jayachandran2014optimal}. The nonlinear part is $\hat{f}  \coloneqq \begin{bmatrix}0, \; \hat{f}_2,\; -\hat{f}_2,\; \hat{f}_4,\; 0, \; 0, \;0,\; 0 \\ \end{bmatrix}^\top$, where \vspace{-2mm}
\begin{align}
    \hat{f}_2(x; \theta) & \coloneqq -\theta_1 x_2(\theta_2 + x_2)^{-1}, \label{fhat2}\\
     \hat{f}_4(x; \theta) & \coloneqq \left(\frac{\theta_3}{1 + \left(x_8/\theta_5 \right)^{\theta_4}}  - \frac{\theta_6  x_3}{\theta_7 + x_3} \right)x_4. \label{fhat4} \vspace{-6mm}
\end{align}
$\hat{f}_2$ describes how 6-MP in the plasma $(x_2)$ is broken down into 6-TGN $(x_3)$. $\hat{f}_4$ describes how the expansion rate of the proliferating white blood cells $(x_4)$ is influenced by the mature white blood cells $(x_8)$ and the drug-derived substance $x_3$. The linear part $A(\theta)x + Bu$ is given by \vspace{-1.5mm}
\begin{equation}
    A(\theta)x + Bu = \begin{bmatrix}-\theta_8 x_1 + u \\ \theta_8 x_1 - \theta_9 x_2 \\ -\theta_{10} x_3 \\ -\theta_{11} x_4 \\ \theta_{11} (x_4 - x_5) \\ \theta_{11} (x_5 -  x_6) \\ \theta_{11} (x_6 -  x_7) \\ \theta_{11} x_7 - \theta_{12} x_8 \end{bmatrix},
\vspace{-1.5mm}\end{equation}
which includes the ingestion of 6-MP and the proliferation and maturation of the white blood cells. 
One can show that $f_t$ \eqref{ftexample} satisfies Parts (a) and (f) of A1. The amounts of mature white blood cells ($x_8$) and neutrophils ($\theta_{13} x_8$) are relevant for clinical decisions. We choose $y \in \mathbb{R}^2$, where $y_1$ and $y_2$ are the measured amounts of mature white blood cells and neutrophils, respectively, and thus, 
\vspace{-3mm}
\begin{equation}
    C(\theta) = \begin{bmatrix}0 & 0 & \cdots & 0 & 1 \\0 & 0 & \cdots & 0 & \theta_{13} \end{bmatrix} \in \mathbb{R}^{2 \times 8}, \vspace{-3mm}
\end{equation}
where $\theta_{13} \in (0,1)$ is the ratio between neutrophils and mature white blood cells (Table \ref{statestable}). The measurement equation $h_t$ \eqref{htexample} satisfies Parts (d) and (f) of A1.

Lastly, we specify cost functions $\hat{c}_t$ to penalize (i) deviations of the amount of neutrophils with respect to a desired range $[\underline{b}, \bar{b}]$ and 
(ii) low doses of 6-MP. The first criterion serves to protect the patient's immune system, while the second criterion serves to limit the production of cancer cells. For example, we may choose \vspace{-1.5mm}
\begin{subequations}\label{myex}
\begin{align}
     \hat{c}_t(x,u;\theta) & \coloneqq \textstyle \frac{1}{N+1}\big( \zeta(x; \theta) - \hat{\lambda}u^2 \big), \quad t \in \mathbb{T}, \\
     \hat{c}_N(x;\theta) & \coloneqq \textstyle \frac{1}{N+1}\zeta(x; \theta), \\
     \zeta(x; \theta) & \coloneqq (\theta_{13} x_8 - \underline{b})(\theta_{13} x_8 - \bar{b}), \label{myzeta}
     \vspace{-3mm}
\end{align}
\end{subequations}
with $\hat{\lambda} \in \mathbb{R}_+$, where $\zeta$ penalizes deviations of the amount of neutrophils with respect to $[\underline{b}, \bar{b}]$. $\hat{c}_t$ \eqref{myex} is continuous and hence l.s.c. for every $t \in \mathbb{T}_N$. $\zeta$ is b.b. since it is convex quadratic. Since $0 \leq u \leq \bar{u}$, we have that $-\hat{\lambda}u^2 \geq -\hat{\lambda}\bar{u}^2$. Thus, $\hat{c}_0,\hat{c}_1,\dots,\hat{c}_N$ \eqref{myex} satisfy Part (b) of A1.
\vspace{-2mm}

\section{Conclusion}\vspace{-3mm}
We have provided a theoretical foundation that unifies several features of biomedical systems, e.g., multiple time scales, partial observability, and uncertain parameters.
A future theoretical step is to analyze the performance of the control policy from Theorem \ref{thmpolicy} under the true dynamics \eqref{gensys}, which depend on $\theta$ rather than $\hat{\theta}_t$.
A shortcoming of Section~\ref{example} is that the partial observability and high dimensionality of the model (which is based on models from \cite{jayachandran2015model} and \cite{jayachandran2014optimal}) preclude the practical application of Theorem \ref{thmpolicy}. Indeed, POMDPs are notorious for their numerical complexity, except in special cases such as the linear-quadratic-Gaussian case, which does not apply here. Hence, one aim of our ongoing work is  the development of methodology with reduced complexity.
We are collecting leukemia patient data, which can be useful for reducing the parameter space (i.e., use the data to identify the subset of parameters that are common to the patient population, as in, e.g., \cite{jost2020modelfrontiers}). Taking inspiration from the framework proposed here, we hope to devise new methods that are resilient to process and measurement noise but also are computationally tractable, which may be facilitated by patient data and domain-specific modeling structure.
\vspace{-2mm}

In the long term, we envision a technology that provides optimized doses for the next cycle 
given the patient's prior doses and measurements.
During a clinical visit,
the patient and her oncologist would see on a computer screen how the 
predicted trajectories of mature white blood cells and neutrophils refit to the patient's data set when a new measurement becomes available. The optimized doses would be compared to the typical doses visually, and then the oncologist would decide which doses to prescribe for the next cycle. The technology may benefit from the theoretical foundation that we have developed here to conceptually unify the unobservable state, unknown parameters, multiple time scales, and nonlinear stochastic dynamics. \vspace{-4mm} 

\begin{ack}           \vspace{-3mm}                    
M.P.C. appreciates Chuanning Wei, Zhengmin Yang, Zehua Li, and Huizhen Yu for fruitful discussions, and she appreciates support provided by the Edward S. Rogers Sr. Department of Electrical and Computer Engineering, University of Toronto. The authors wish to thank two anonymous reviewers for their helpful suggestions. \vspace{-2mm} 
\end{ack}

\bibliographystyle{plain}        
\bibliography{autosam}           
                                        
\end{document}